\documentclass[entropy,article,accept,pdftex,moreauthors]{Definitions/mdpi}
\firstpage{1}
\pubvolume{1}
\issuenum{1}
\articlenumber{0}
\pubyear{2025}
\copyrightyear{2025}
\externaleditor{Ángel S. Sanz} 
\datereceived{31 October 2025}
\daterevised{24 November 2025} 
\dateaccepted{1 December 2025}
\datepublished{ }
\hreflink{https://doi.org/} 

\usepackage{bbold}

\Title{Optimal Transfer of Entanglement in Oscillator Chains in Non-Markovian Open Systems}

\TitleCitation{Optimal Transfer of Entanglement in Oscillator Chains in Non-Markovian Open Systems}

\Author{{Da-Wei} Luo *\orcidA{}, Edward Yu and Ting Yu}

\AuthorNames{Da-Wei Luo, Edward Yu and Ting Yu}

\isAPAStyle{%
       \AuthorCitation{Luo, D.W., Yu, E., \& Yu, T.}
         }{%
        \isChicagoStyle{%
        \AuthorCitation{Luo, Da-Wei, Edward Yu, and Ting Yu.}
        }{
        \AuthorCitation{{Luo,}  D.-W.; Yu, E.; Yu, T.}
        }
}

\address[1]{%
Center for Quantum Science and Engineering{,} 
 Department of Physics, Stevens Institute of Technology, \mbox{Hoboken, NJ 07030, USA;} {eyu2@stevens.edu (E.Y.); tyu1@stevens.edu (T.Y.)}}

\corres{\hangafter=1 \hangindent=1.05em \hspace{-0.82em}Correspondence: dluo2@stevens.edu}

\abstract{
We consider the transfer of continuous-variable entangled states in coupled oscillator chains embedded in a generic environment. We demonstrate high-fidelity transfer via optimal control in two configurations---a linear chain and an X-shaped chain. More specifically, we use the Krotov optimization algorithm to design control fields that achieve the desired state transfer. Under environmental memory effects, the Krotov algorithm needs to be modified, since the dissipative terms in non-Markovian dynamics are generally governed by the time-dependent system Hamiltonian. Remarkably, we can achieve high-fidelity transfer by simply tuning the frequencies of the oscillators while keeping the coupling strength constant, even in the presence of open-system effects. For the system under consideration, we find that quantum memory effects can aid in the transfer of entanglement and show improvement over the memoryless case. In addition, it is possible to target a range of entangled states, making it unnecessary to know the parameters of the initial \mbox{state beforehand.}}


\begin{document}

\section{Introduction}

Quantum entanglement~\cite{Horodecki2009a} is a unique phenomenon of quantum mechanics and~is one of the most important resources for quantum technologies~\cite{Nielsen2000a}. It lies at the heart of various quantum information and computation tasks, such as teleportation~\cite{Zeilinger2000a,Pirandola2015a}, quantum key distribution~\cite{Ekert1991a}, and~the Grover search algorithm~\cite{Grover1996a}, to name but a few. Recent advances in quantum-enhanced sensing and metrology technologies~\cite{Degen2017a, DeMille2024a,Lecamwasam2024a,Wang2024a,Agarwal2022a,Weiss2021a,Giovannetti2004a} have also shed light on how entanglement can be used to implement high-precision quantum metrology devices, where the use of the entangled state~\cite{Demkowicz-Dobrzaifmmode-nelse-nfiski2014a,Yang2024a,Lecamwasam2024a,Wang2024a,Xia2023a,Dooley2023a} has been shown to push measurement precision to the Heisenberg limit~\cite{Huang2024a}.
While discrete quantum systems are useful as implementations for qubits in quantum information and computation tasks, in~quantum metrology tasks continuous-variable systems~\cite{Kwon2022a,Fadel2024a,Sun2020a} have enjoyed wide use, where different quantum optical systems~\cite{Dowling2008a,Dowling2015a,Huver2008a} or hybrid opto-mechanical systems~\cite{Aspelmeyer2014a,Xia2023a,Li2021a,Zobenica2017a} are some prominent~examples.

In this paper, we study the optimal control of quantum entanglement in continuous-variable systems and~how to adapt the control in non-Markovian open systems. Various cutting-edge strategies for the optimal control of entangled states or quantum dynamics in general have recently been proposed, such as an adaptive procedure utilizing analytical Lie algebraic derivatives~\cite{Goodwin2022a} and Lie group and algebraic control for high-dimensional qudits~\cite{Omanakuttan2023a}, where control may be realized by intertwining a sequence of local SU gates and the interaction in alternating layers of single-qudit gates and entangling gates. Another interesting approach uses the adiabatic invariant in Su--Schrieffer--Heeger chains~\cite{Palaiodimopoulos2021a}, where one may take advantage of resonant effects to speed up the control. More recently, various machine learning techniques~\cite{Liu2024a,Duncan2025a,Xie2022a} such as reinforcement learning have also been applied to derive quantum control, which has been shown to be scalable to large systems. Here, we will focus on the use of a gradient-based iterative algorithm known as Krotov's method~\cite{Konnov1999b,Reich2012v,Tannor1992h, Jager2014a,Goerz2019a}, which allows us to design complex control targets and also lets us fully consider the influences of the non-Markovian effects. We study two configurations of quantum harmonic oscillator chains and~apply optimal control to transfer entanglement states from one end of the chain to the other. We show how to revise the optimal control for non-Markovian open systems, which introduces dissipative terms that are dependent on the external controls. The~techniques used here are designed to be generic and can be applicable to other discrete- or continuous-variable~systems.

When a system of interest is coupled with an external environment,  the~system generally needs to take environment noises and other decoherence factors into consideration. This more precise description of quantum dynamics is studied in the framework of open quantum systems~\cite{Breuer2002,Rivas2012a}, where quantum systems no longer evolve unitarily and can display dissipative behaviors.  Quantum open systems model the system under consideration with its surroundings as a composite system evolving according to the system-plus-bath Hamiltonian, where the system dynamics can be extracted by tracing out the environment degrees of freedom. Under~the influences of open-system effects, quantum entanglement can display some intricate behaviors, such as sudden death and births~\cite{Yu2009a,Yu2006a}. Analytically, this can make it a challenging task to derive the dynamical equations for the system, 
and~one common approximation is to consider a flat-spectrum bath (white noise) and discard the memory effects of the environment. This is known as the Markov approximation~\cite{Lindblad1976a,Breuer2002}. However, such an approximation cannot track the backflow of information from the environment to the system and~can fail to be applicable when the environment is structured or when the system--environment coupling strength is strong. In~such cases, one needs to consider the full non-Markovian~\cite{Chin2012a,Yu1999a,Diosi1998a,Vega2017a,Breuer2016a} effects. While analytically more complicated, non-Markovian dynamics describes the system dynamics in a more precise fashion and~can account for various interesting physical phenomena. For~example, it has been shown to aid in the production or preservation of quantum entanglement~\cite{Zhao2011a} and enhance certain quantum algorithms~\cite{Li2020a} and~quantum metrology tasks~\cite{Chin2012a,Yang2024b}.

 This paper is organized as follows. We first introduce the quantum systems under consideration and control strategy using Krotov's method. We then present the details on how to adapt Krotov's method to non-Markovian open systems, and a~comparison with the Markov case is also made. Practical considerations such as how to limit the control amplitudes and unknown state parameters are also considered. Mathematical details are left to~{Appendices} \ref{appx_opencm_vec} and \ref{appx_krotovo}.

\section{Optimal Transfer of Entangled States in Oscillator Chains: Closed-System~Setup} \label{sec_sys}

{For} modeling a chain of harmonic oscillators, we consider a quadratic Hamiltonian of the {form}
\begin{align}
 H_0 = \frac{\omega_0}{2} \sum_j \left( p_j^2 + q_j^2 \right) + \sum_{j,k} g_{j,k} \eta_j^\dagger \eta_k
\end{align}
where \(p_j, q_j\) are the canonical momentum and position operators,\vspace{12pt}
\begin{align}
 p_j = \frac{i}{\sqrt{2}} \left( a_j^\dagger - a_j \right), \quad
 q_j = \frac{1}{\sqrt{2}} \left( a_j^\dagger + a_j \right),
\end{align}
\(a_j\) is the annihilation operator for the \(j\)-th oscillators, and~\(\eta_i\) determines the coupling mechanism: \(\eta_i = a_i\) for a rotating-wave approximation-like coupling that preserves the total number of excitations and~\(\eta_i=q_i\) for position--position~coupling.

\begin{figure}[H]
 \includegraphics[width=.9\textwidth]{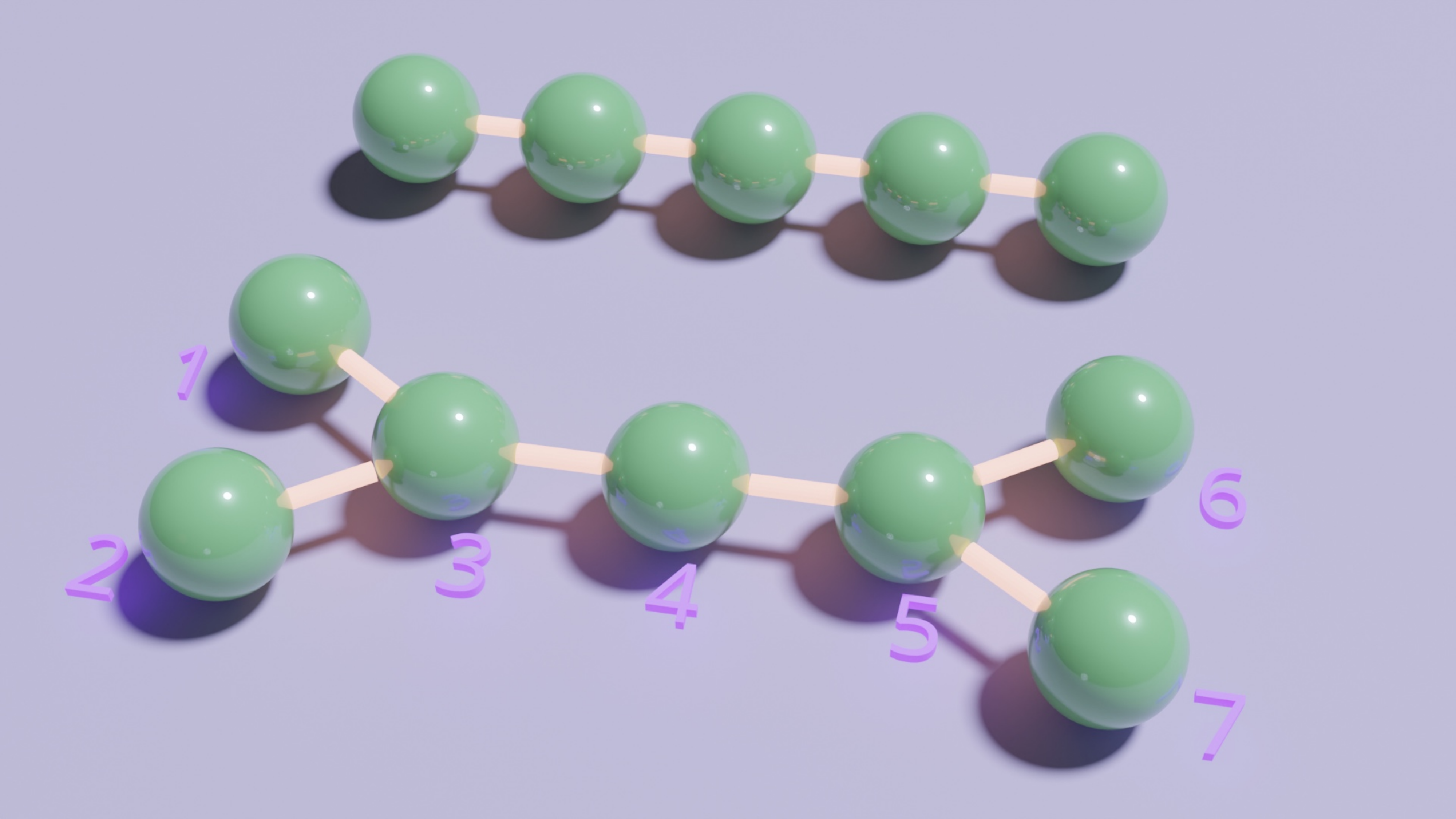}
 \caption{{Schematic} of the models under consideration. We have considered two types of oscillator chains, a linear chain and an X-shaped chain, with~the goal of transferring entangled states through the coupled~chains.}\label{fig_models}
\end{figure}

The control field would be applied to tune only the frequencies of the oscillators,
\begin{align}
 H_{c, i} = \frac{1}{2} \left( p_i^2+q_i^2 \right),
\end{align}
and the total Hamiltonian of the system is then \( H_{s}(t) = H_0 + \sum_i c_i(t) H_{c, i}(t) \). This setup may allow for an easier experimental realization, since the coupling strengths between the harmonic oscillators could be difficult to control in a real-time manner~\cite{Milne2020a,Bishof2013a}. Experimentally, it may be realized as a chain of superconducting quantum interference devices (SQUIDs) or circuit quantum electrodynamics systems
as a chain of LC circuits~\cite{Blais2007a,Blais2021a}, where individual frequency tuning may be possible~\cite{Liao2010a} by either controlling the boundary condition of the electromagnetic wave in a transmission line~\cite{Wallquist2006a} or~changing the effective inductance in the circuit quantum electrodynamics model~\cite{Castellanos-Beltran2007a}. Here, we consider two configurations of the coupled chains, a~simple linear chain and~an X-shaped chain connecting two oscillators at two ends (schematically shown in Figure~\ref{fig_models}).

For this many-state system, due to the exponential growth of the Hilbert space, even with a modest cut-off \(N_c\) of the Fock state basis, the~dimension of the Hilbert space would be \(N_c^N\) for a chain with length \(N\), and directly solving this system numerically can be a challenging task. On~the other hand, it should be noted that all Gaussian states can be uniquely determined by their first and second moments, 
encoded in the expectation values \(v_i = \langle R_i \rangle\) and the covariance matrix (CM) \(\gamma_{i,j} = \langle \left\{R_i R_j \right\} - 2 \langle R_i \rangle \langle R_j \rangle\rangle\), where \(R = \left[q_1, q_2, \ldots, q_N, p_1, p_2, \ldots, p_N\right]\) is the canonical position and momentum operators for each site of the chain. The~canonical commutation relationship is given as a symplectic form \(\left[R_i, R_j\right] = i \sigma_{i,j}\),
\begin{equation}
  \sigma = \begin{bmatrix}
    0_N & I_N\\ -I_N & 0_N
  \end{bmatrix},
\end{equation}
where \(I_N\) (\(0_N\)) are identity (zero) matrices of size \(N\times N\). Under~quadratic Hamiltonians \(H = R M R^T/2\), Gaussian states will remain Gaussian and follow the equation of motion
\begin{align}
  \partial \gamma = \sigma \bar{M} \gamma + \gamma \left[\sigma \bar{M}\right]^T, \label{eq_dcm_clsd}
\end{align}
in closed systems, where we have the symmetrized \(\bar{M}(t) = (M+M^T)/2\).

The quantum state transfer and dynamics of entanglement of oscillator chains have been studied in many different situations, such as using a translation-invariant chain without control~\cite{Plenio2005b} or~tailoring laser field pulses of cascading systems~\cite{Parkins1999a}. For~the two configurations described above, we will show how to  utilize the quantum optimization controls~\cite{Halaski2024a,wiseman2009quantum,gough2012principles,Jager2014a,Genoni2013a,bonnard2012optimal,Doria2011a,Muller2022a,Konnov1999b,Reich2012v,Gollub2008a,sklarz2002loading,morzhin2019krotov,Goerz2019a,miki2023generating,cong2014control,brif2010control,Rojan2014a} to achieve the transfer of the entangled state in the presence of environmental noises.
 A key aspect of the optimal control method is the construction of an appropriate optimization functional \(J\)
to be minimized. This functional typically includes a figure of merit, such as the fidelity for state preparation. Once  \(J\)
 is defined, an~optimization algorithm is selected to determine the control functions that minimize it. Gradient-free algorithms generally converge more slowly, except~when the number of optimization parameters is small. In~contrast, gradient-based methods require the computation of the derivative of the optimization functional, which can be obtained either analytically or numerically through automatic differentiation~\cite{Goerz2022a,zygote_jl}.
To date, various optimization control algorithms have been proposed, such as the stimulated Raman adiabatic passage (STIRAP)~\cite{Vitanov2017a}, Gradient Ascent Pulse Engineering (GRAPE)~\cite{Khaneja2005a}, and~gradient-free chopped random-basis quantum optimization~\cite{Doria2011a,Muller2022a} methods. Optimal controls using machine learning tools have also been recently proposed~\cite{Goerz2022a,Niu2019a,Sivak2022a}. In~this work, we employ Krotov's method~\cite{Konnov1999b,Reich2012v,Tannor1992h, Jager2014a,Goerz2019a,Chen2025a,Yu2023a}, which is an iterative, gradient-based algorithm. Through a clever separation of the interdependence of the quantum state and the control field, it takes in an initial guess control and iteratively updates the controls such that the optimization functional is guaranteed to be monotonically decreasing: denote the optimization functional at iteration \(k\) as \(J^{(k)}\); we have \(J^{(k+1)} < J^{(k)}, \, \forall k\).
The optimization target functional in Krotov's method is taken to be
\begin{align}
  J^{(i)}\left[|\varphi^{(i)}(T) \rangle, \{c_l^{(i)}(t)\}\right]
    = J_T(|\varphi^{(i)}(T) \rangle) + \sum_l \int dt g(c_l^{(i)}(t)), \label{eq_ctrlj}
\end{align}
where $|\varphi^{(i)}(t) \rangle$ is the wave functions at the $i$-th iteration at time $t$, evolving under the controls $c_l^{(i)}$ of the $i$-th iteration, following any Schr\"odinger-like equation
\begin{equation}
  \partial_t |\varphi^{(i)}(t) \rangle = -i\left[ H_0 + \sum_k c_k^{(i)}(t) H_{c, k}(t)\right] |\varphi^{(i)}(t) \rangle,
\end{equation}
Note here that we do not require the Hamiltonians to be Hermitian---any linear, homogeneous differential equation of the form \(\partial_t V = -iW_{\rm eff}V\) may be considered for~vector $V$ and matrix $W_{\rm eff}$. $J_T$ is a final time objective function to minimize and $g$ is a correction term of the running cost of the control fields, usually taking the form of
\begin{align}
 g( c^{(i)}_l(t) ) = \frac{\Lambda_{a,l}}{S_l(t)}(\Delta c_l^{(i)}(t))^2,
\end{align}
where $\Lambda_{a,l}>0$ is an inverse step-size, $\Delta c_l^{(i)}(t)= c_l^{(i)}(t) - c_l^{(i-1)}(t)$ is the control function update between the current and last iteration, and~$S_l(t) \in [0,1]$ is an update shape function, generally taken as the Blackman window function~\cite{Ohtsuki2003a,Goerz2019a}. The~control pulse can then be updated iteratively using
\begin{align}
 \Delta c_l^{(i)}(t) = \frac{S_l(t)}{\Lambda_{a,l}} \mathrm{Im} \left[\left\langle \chi^{(i-1)}(t)\left|\frac{\partial H^{(i)}}{\partial c^{(i)}_l(t)} \right| \varphi^{(i)}(t)\right\rangle\right], \label{eq_ctrlup}
\end{align}
where $| \chi^{(i)}(t) \rangle $ is a co-state that evolves `backwards' according to $H^\dagger(t)$, with~the boundary condition at the final $T$ as $|\chi^{(i-1)}(T) \rangle = - \partial J_T/\partial \langle \varphi^{(i-1)} (T)|$. By~construction, the~Krotov control ensures the monotonic convergence of the iterative algorithm in that the control objective function Equation~\eqref{eq_ctrlj} of the current iteration is guaranteed to be smaller than that of the previous iteration. 
 We column-stack the CM \( \vec{\gamma} \), and~using the Kronecker-product trick \(AOB = [B^T \otimes A] \vec{O}\) we can cast Equation~\eqref{eq_dcm_clsd} into the homogeneous linear differential equation form required by Krotov's method,
\begin{align}
  \partial_t \vec{\gamma} = [I \otimes \sigma \bar{M} + \sigma \bar{M} \otimes I] \vec{\gamma}
\end{align}

The entangled state to be transferred can be chosen as a two-mode squeezed state (TMSS)~\cite{Ekert1989a,Agarwal2012a}, \(S_{i,j}(r) |0 \rangle_{1\ldots N}\), where \(S_{i,j}(r)\) is the two-mode squeezing operator acting on mode \(i, j\),
\begin{equation}
  S_{i,j}(r) = \exp\left[ r^* a_i a_j - r a_i^\dagger a_j^\dagger \right].
\end{equation}
The initial state would then be \(|\Psi_i(r) \rangle = S_{1,2}(r) |0 \rangle_{1\ldots N}\), with~a target state \(|\Psi_T(r) \rangle = S_{N-1,N}(r) |0 \rangle_{1\ldots N}\). As~the first example, we consider a linear chain of length \(N=5\), with~nearest neighbor coupling that preserves the excitation \(\eta_j = a_j,\, g_{i, i+1} = g_{i+1, i} = g_0\)
\begin{align}
  H_0 = \omega_0 \sum_i a_i^\dagger a_i + g_0 \sum_{i=1}^{N-1} a_i^\dagger a_{i+1} + h.c.
\end{align}
One subtle but important point in choosing the control's optimization functional is that while it may seem natural to choose fidelity as the figure of merit, we should note that fidelity only measures the overlap between states and~is not necessarily a good indicator of the entanglement information. That is, two quantum states can have high overlap in terms of fidelity but~different measured entanglement degrees.  Strictly speaking  this is 
only guaranteed if fidelity is exactly \(1\) without considering the entanglement content. In~this paper, for~the control target optimization functional \(J_T\), we compare the minimization of two non-negative and normalized functionals: \(J_1 = F_r\) and \(J_2 = (F_r + N_r) / 2\), where \(F_r, N_r\) are the normalized residuals of the fidelity and entanglement measured by the logarithmic negativity~\cite{Vidal2002a,Horodecki2009a,Simon2000a,Plenio2005a}, respectively,
\begin{align}
F_r = 1 - F,\quad N_r = \left( \frac{N - N_0}{N+N_0} \right)^2
\end{align}
where the fidelity~\cite{Uhlmann1976a} \(F\) between two continuous variable states may be expressed~\cite{Banchi2015a} using the two CMs \(\gamma_1, \, \gamma_2\) as
\begin{align}
  F &= \frac{F_{\rm tot}}{\sqrt[4]{\det(\gamma_1/2, \gamma_2/2)}}, \quad F_{\rm tot} = \prod \left[w_k + \sqrt{w_k^2-1}\right]^{1/2}, \label{eq_cv_fid}
\end{align}
since here the state we chose has \(v_i = \langle R_i \rangle = 0\),
where \(w_k\) are the eigenvalues of the auxiliary matrix \(W=-2V_{\rm aux}i \sigma\) and
\begin{equation}
  V_{\rm aux} = \sigma^T \left(\gamma_1/2 + \gamma_2/2\right)^{-1} \left( \sigma + \gamma_2 \sigma \gamma_1 \right) / 4.
\end{equation}
Note that the fidelity is taken to be the Bures fidelity \(\mathcal{F} = \operatorname{Tr}\sqrt{\sqrt{\rho_1}\rho_2\sqrt{\rho_1}}\), which differs from the Uhlmann--Jozsa fidelity by \(F_{UJ}=\mathcal{F}^2\).
The logarithmic negativity can be \mbox{obtained with}
\begin{align}
	N = -\sum_i \log_2\left(\min(1, |\lambda_i|)\right) \label{eq_negdef}
\end{align}
where $\lambda_i$ are the symplectic eigenvalues (accounting for the 2-fold degeneracy) of the partial-transpose CM $\gamma^{T_B} = P \gamma P$, $P=\mathrm{diag}(1,1,1,-1)$~\cite{Simon2000a,Plenio2004a}. While obtaining the analytical derivative of the continuous-variable fidelity Equation~\eqref{eq_cv_fid} or negativity Equation~(\ref{eq_negdef}) is a hard task, it may be easily obtained numerically without approximation using an automatic differentiation technique~\cite{Goerz2022a,zygote_jl,Innes2019a}. 

\begin{figure}[H]
  \hspace{-10pt}
  \includegraphics[width=.9\textwidth]{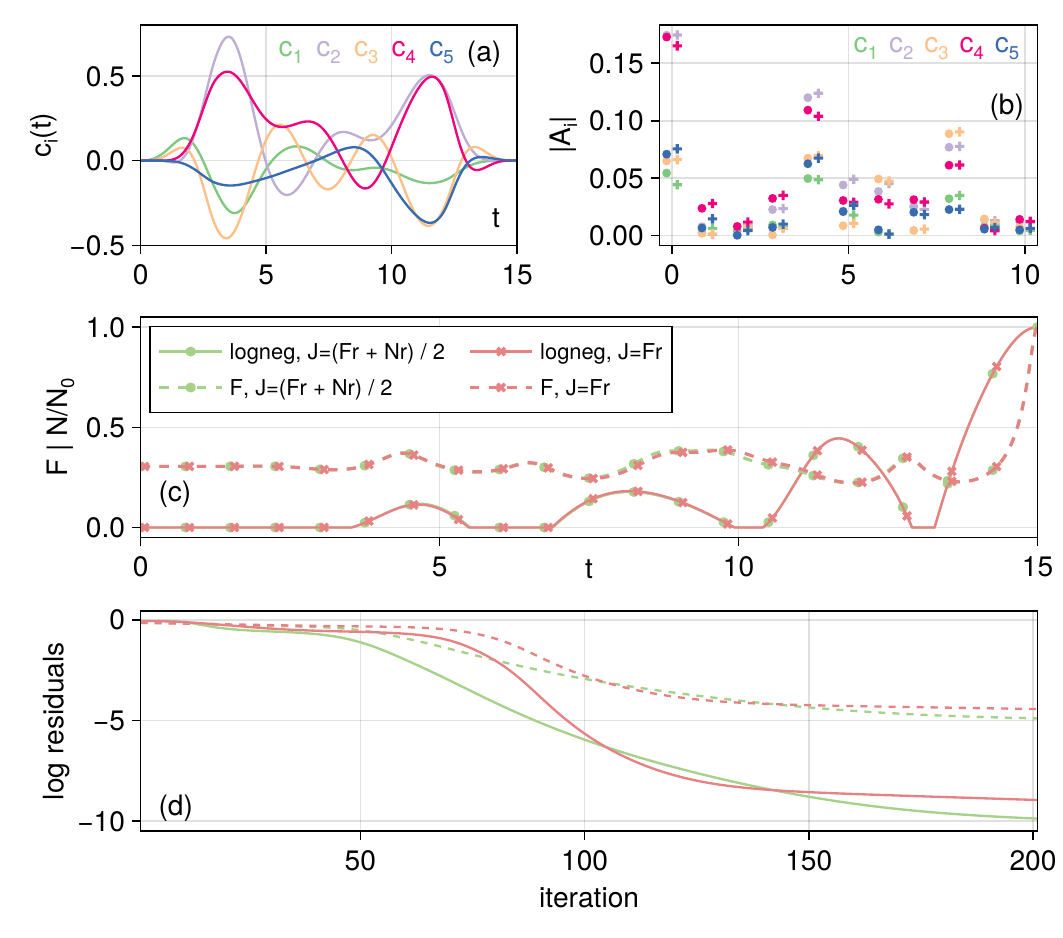}
 \caption{\textls[3]{{Transferring}
  the entangled state through a linear chain as a closed system. Panel (\textbf{a}): The control fields applied as functions of time, with the control targeting both fidelity and negativity. Different colors correspond to the control field applied to each oscillator. Panel (\textbf{b}): \textls[20]{First 10 discrete Fourier transform frequencies of controls targeting just the fidelity (circle markers) and controls}}}\label{fig_lch_fnc}
\end{figure}

Taking \(\omega_0 =1\), \(g_0 = 0.4\), a~total runtime \(T=15\), and~squeezed parameter \(r = 1.2\), we can now carry out the optimization for the linear chain. The~initial guess fields are just set to a constant \(c_i = 0\).
We show the controls \(c_i\) under the target optimization functional \(J_2\) as functions of time in Figure~\ref{fig_lch_fnc}a. We show that these functions are well-behaved; that is, they neither grow unbounded nor exhibit rapid oscillations. These properties form the basis for potential experimental realizations.
The first 10 discrete Fourier transform frequencies on a grid size of \(2000\) are plotted in {Figure}~\ref{fig_lch_fnc}b, where we can see that the two optimization functionals lead to slightly different control fields, with~similar dynamics of the fidelity and negativity shown in  {Figure}~\ref{fig_lch_fnc}c. We show the residuals \(F_r, N_r\) in {Figure}~\ref{fig_lch_fnc}d, and~we can see that targeting both fidelity and negativity (green lines) leads to smaller residuals, illustrating that the entangled state transfer has been achieved at high fidelity while also ensuring the target negativity value has also been~reached.\vspace{-6pt}

{\noindent{}\small{targeting both fidelity and negativity ($+$markers). Panel (\textbf{c}): Fidelity and negativity dynamics under the two controls, where negativity is normalized with respect to the target value. Panel (\textbf{d}): Logarithmic (base 10) of the residuals of the target function as a function of Krotov control iterations. Panels (\textbf{c},\textbf{d}) share the same legends.
}}
\vspace{12pt} 

\begin{figure}[H]
 \includegraphics[width=.9\textwidth]{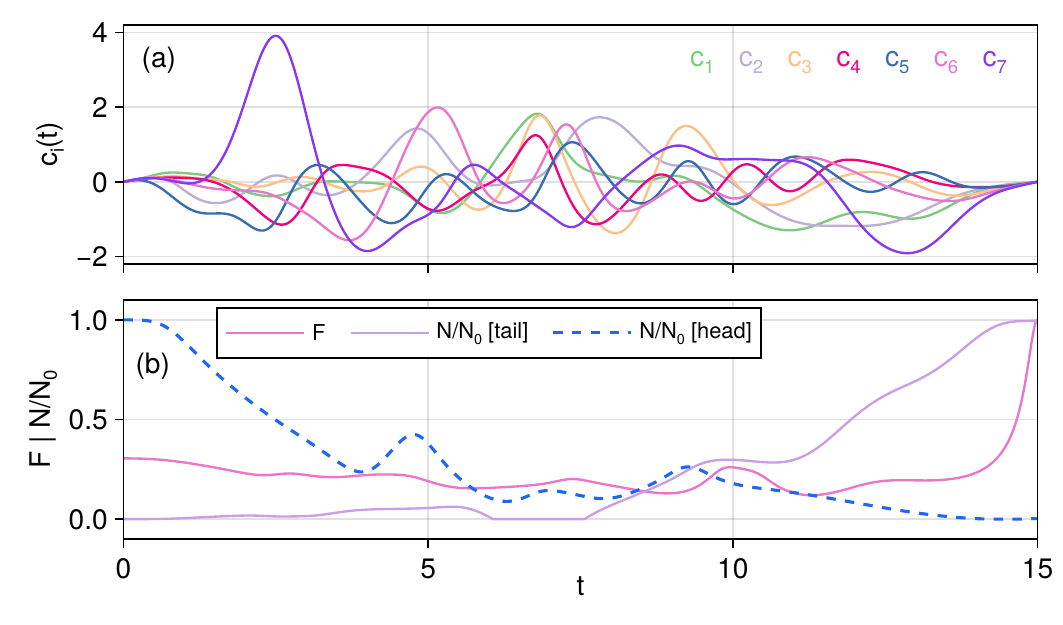}
 \caption{Transferring the entangled state through an X-shaped chain as a closed system. Panel (\textbf{a}): Control fields as functions of time. Panel (\textbf{b}): Fidelity and negativity dynamics under control, illustrating the transfer of entanglement from the head of the chain to the~tail.}\label{fig_x_closed}
\end{figure}

Next, we consider an X-shaped chain with length \(N=7\), with \(\eta_i\) set to position--position neighboring couplings, where the Hamiltonian is given by
\begin{align}
  H_0 = \frac{1}{2}\sum_i \left[p_i^2 + q_i^2\right] + g_0 \left(q_1 q_3 + q_2 q_3 + \sum_{i=3,4} q_i q_{i+1} + q_5 q_6 + q_5 q_7 \right)
\end{align}
The initial guesses are taken to be simple sine functions, \(c_i = (0.1 + i / 20) \sin(4\pi t/T) \). The~resulting control fields and controlled dynamics are shown in Figure~\ref{fig_x_closed}. While entanglement is not a conserved quantity, it is nevertheless still interesting to observe that during the entangled state transfer, the~entanglement between sites 1 and 2 decreases while the entanglement at the tail end gradually increase to the~target.

\section{Krotov's Method in Open Quantum~Systems} \label{sec_nmc}

In realistic scenarios, all quantum systems inevitably interact with their surrounding environment, and~the resulting open-system noise typically leads to decoherence and the degradation of the system's quantumness. There are two main approaches to studying optimal control in the context of quantum open systems. The~first is to apply control fields originally designed for closed systems and evaluate their robustness against environmental noise. The~second is to design new control fields that explicitly account for the open-system dynamics, treating the dissipative part as the uncontrolled evolution. 
In~this paper, for~simplicity, we consider a zero-temperature bosonic environment.
\begin{equation}
  H_{\rm tot} = H_s + H_b + H_{\rm int} = H_s + \sum_k \left( \tilde{\omega}_k b_k ^\dagger b_k + g_k L^\dagger b_k ^\dagger + g_k^* L b_k \right),
\end{equation}
where \(b_k\) is the annihilation operator of the \(k\)-th bath mode with frequency \(\tilde{\omega}_k\), \(g_k\) is the coupling strength, and \(L\) is the system--bath coupling~operator.

For open-system dynamics, the~widely used Markov approximation is valid when the environmental bath is structureless and the system–environment coupling is sufficiently weak. However, such conditions are often not met in realistic settings, and~a more accurate description of the system's evolution must account for non-Markovian memory effects. To~derive the equations of motion governing non-Markovian open-system dynamics, we employ the quantum state diffusion (QSD) formalism to first obtain a corresponding non-Markovian master equation~\cite{strunz1999open,Yu1999a,qsd_n1}. The~QSD equations project the bath modes onto coherent states and lead to a set of stochastic trajectories
\begin{equation}
	\partial_t |\psi_t(z^\ast)\rangle
	=\bigg[-iH_s + Lz^*_t - L^\dagger \bar{O}(t,z^*)  \bigg]|\psi_t(z_t^\ast)\rangle,
\end{equation}
where
\begin{equation}
  O(t,s,z^*)\psi_t\equiv\frac{\delta\psi_t}{\delta z_s}
\end{equation}
is an ansatz operator for the functional derivative with the initial condition $O(t, s=t, z^\ast) = L$ and $\bar{O}(t,z^\ast)\equiv\int^t_0ds\alpha(t,s)O(t,s,z^*)$. The~reduced density operator may be obtained by a stochastic average $\rho = M[| \psi_t(z_t^*) \rangle \langle  \psi_t(z_t) |]$, where $M[\cdot]\equiv\int\frac{dz^2}{\pi}e^{-|z|^2}[\cdot]$ represents the average over the noises.
The noise here is chosen to be Ornstein--Uhlenbeck noise, which corresponds to a Lorentzian bath spectrum with correlation function
\begin{equation}
	\alpha(t,s)=\frac{\xi}{2}e^{-(\xi+i\Omega)|t-s|}, \label{eq_alphats}
\end{equation}
where $1/\xi$ represents the memory time and $\Omega$ signifies a central frequency shift. This choice of the correlation function allows us to study how the system behaves under a non-Markovian bath with a continuously tunable strength of the memory effects, with~smaller $\xi$ corresponding to stronger memory effects, whereas $\xi \rightarrow \infty$ would lead to a memoryless Markov dynamics. In~principle, other types of correlation functions may be expanded~\mbox{\cite{Hartmann2017a,Ritschel2014a}} as a linear combination of Equation~\eqref{eq_alphats}, and~finite-temperature baths may be cast as a fictitious thermal--vacuum state~\cite{Yu2004a}, so the calculations carried out here may be extended to other types of spectra.
The $O$-operator follows the consistency condition
\begin{equation}
  \partial_t \frac{\delta}{\delta z_s} | \psi_t(z_t^*) \rangle = \frac{\delta}{\delta z_s} \partial_t | \psi_t(z_t^*) \rangle.
\end{equation}
The \(O\)-operator has been analytically obtained for a wide range of interesting models~\mbox{\cite{Jing2013a,Jing2010a,Strunz2004a}} and~can be numerically calculated up to arbitrary order~\cite{Luo2015a,Li2014a} by an expansion of different orders of noise terms. It is also worth pointing out that in most cases, even the noiseless leading order suffices to capture the interesting non-Markovian effects. In particular, in~this case the master equation is readily given by
\begin{align}
	\frac{\partial}{\partial t} \rho_s(t)&=-i \left[H_s(t),\rho_s(t)\right]+\left[L,\rho_s(t)\bar{O}^{(0)\dagger}(t)\right]-\left[L ^\dagger,\bar{O}^{(0)}(t)\rho_s(t)\right], \label{eq_rho_meq}
\end{align}
where \(\bar{O}^{(0)}(t)\) is the leading order approximation following
\begin{align}
	\partial_t \bar{O}^{(0)}(t) &= \alpha(0)L - \xi_{\rm eff}\bar{O}^{(0)}(t) + \left[-iH_s(t) - L ^\dagger \bar{O}^{(0)}(t), \bar{O}^{(0)}(t)\right]. \label{eq_dodt}
\end{align}
where \(\xi_{\rm eff} = \xi + i \Omega\). In~the Markov limit, we have \(\bar{O}^{(0)}(t) \rightarrow L/2\), and we would recover the usual Lindblad master~equation.

\begin{figure}[H]
 \includegraphics[width=.95\textwidth]{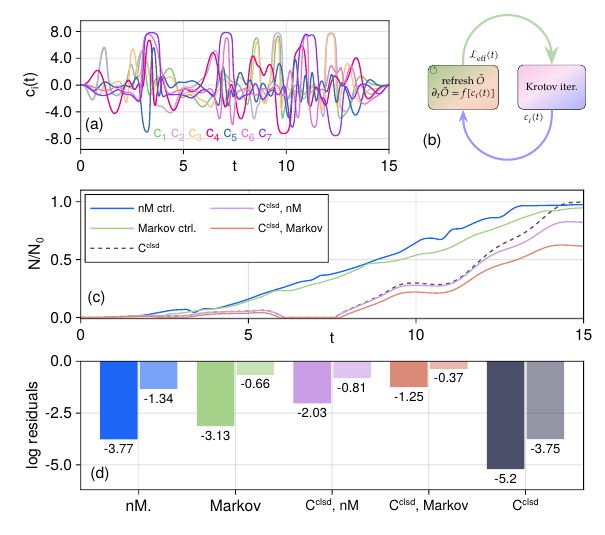}
 \caption{{Optimized} control under open-system effects. Panel (\textbf{a}): Control fields for the non-Markovian open system. Panel (\textbf{b}): Modified Krotov control iteration for non-Markovian open-system dynamics. Panel (\textbf{c}): Entanglement dynamics under different scenarios: deriving a new control under non-Markovian open-system effects (blue solid line), deriving new a control under Markov noise (green solid line), and~applying the closed-system controls in (non-)Markovian open systems as red (purple) solid lines. The~ideal closed-system dynamics is also shown as a black dashed line for reference. Panel (\textbf{d}): Logarithmic of the final residuals of the target function; left solid ones are for the negativity and right fainter ones are for the~fidelity.}\label{fig_nmc}
\end{figure}

It is now clear that extending the Krotov method to Markovian dynamics is relatively straightforward, as~one can simply treat the Lindblad equation as the uncontrolled system. However, the~non-Markovian case requires more careful treatment: the \(\bar{O}^{(0)}\) operator appearing in the dissipative term is governed by the time-dependent system Hamiltonian. Consequently, after~each control field update in the Krotov iterations,
the \(\bar{O}^{(0)}\) operator needs to be re-calculated (Figure~\ref{fig_nmc}b).

Consider a system--bath coupling operator that is linear in the canonical position and momentum, \(L=l_iR_i\); we may write down an ansatz for the \(\bar{O}^{(0)}\)-operator that also only contains linear terms of \(R_k\) as \(\bar{O}^{(0)}(t)=o_i(t) R_i\). 
For~brevity, we drop the explicit time dependence of the \(\bar{O}^{(0)}\) operator~coefficients.

The Langevin equation for the first and second moments may then be readily derived~\cite{Chen2025a}, leading to an equation of motion of the CM,
\begin{align}
	\partial_t \gamma &= \left[\sigma\bar{M} + \sigma \Delta \right] \gamma + \gamma \left[\sigma\bar{M} + \sigma \Delta\right]^T + 2 \left[\sigma \delta^R \sigma^T \right] \label{eq_nm_dcm}
\end{align}
where $\delta_{mn} = l_m^* o_n + o_m^* l_n$, $\delta^R = \mathrm{re} [\delta]$, $\Delta_{mn}=i l_m o_n^* - i l_m^* o_n$, and~we have dropped the explicit time dependence of the operators for brevity. Due to the diffusion term \(2 \left[\sigma \delta^R \sigma^T \right]\) in Equation~\eqref{eq_nm_dcm}, it is not formally a homogeneous differential equation but~can be cast into one (see Appendix~\ref{appx_opencm_vec}) by padding a constant to the column-stacked vector \(\vec{\gamma} \rightarrow \vec{\gamma_1} = [\gamma, 1]\) so it can be formally written as \(\partial_t \vec{\gamma_1}(t)  = \mathcal{L}_{\rm eff}(t)\vec{\gamma_1}(t)\), where \(\mathcal{L}_{\rm eff}(t)\) is a matrix of size \((2N+1) \times (2N+1)\).

A leading-order approximation of the $\bar{O}$ operator that keeps only the noise-independent terms can be shown~\cite{Chen2025a} to follow,
\begin{align}
	\partial_t o_i
	&= \alpha(0) l_i - \xi_{\rm eff} o_i - o_l [\sigma\bar{M}]_{li} - i \sigma_{kl} \left[o_k o_l l^*_i  + o_i o_l l^*_k \right] \label{eq_nm_dodt_inR}
\end{align}

To revise the Krotov iteration for non-Markovian open systems, we need to take the updated Hamiltonian coefficients \(\bar{M}\) after each Krotov iteration and redo Equation~\eqref{eq_nm_dodt_inR} to obtain the correct  \(\mathcal{L}_{\rm eff}(t)\) under non-Markovian noises.
It is also worth pointing out that for the control update, the~contribution of \(\bar{O}^{(0)}\) to \(\partial \mathcal{L}_{\rm eff}/\partial c_i(t)\) may be omitted (see Appendix~\ref{appx_krotovo}) following a perturbative~expansion.

We are now equipped to carry out the optimal control in an open-system setting. Here, we take \(L = \lambda \sum_i q_i\), \(\lambda = 0.3 g_0 \), and~for the non-Markovian bath spectrum we choose the memory parameter \(\xi=0.6\) and central frequency \(\Omega = 0.7\). In~the simulations, the~re-calculation of the \(\bar{O}\) operator is carried out for the first \(100\) iterations and then every \mbox{\(20\) iterations} to make it more efficient. To~prevent the control fields from getting too large in this scenario, we clamp~\cite{Goerz2019a} the controls with a \(\tanh\) function
\begin{align}
  c_i(t) &\rightarrow \tilde{c}_i(t) = A \tanh \frac{c_i(t)}{A} \in [-A, A] \nonumber, \\
  H_c(t) &= \sum_i \tilde{c}_i(t) H_{c, i}.
\end{align}
In this case, the~control update Equation~\eqref{eq_ctrlup} also needs to be revised to follow the \mbox{chain rule,}
\begin{align}
  \frac{\partial H}{\partial c_i(t)} = \frac{\partial \tilde{c}_i(t)}{\partial c_i(t)}  H_{c,i}
  = \operatorname{sech}^2\left[\frac{c_i(t)}{A}\right] H_{c,i}.
\end{align}
Setting \(A=8\), we show the resulting control fields in Figure~\ref{fig_nmc}a, where we can see that the control fields here are well-behaved. The~entanglement dynamics and the logarithmic of the final residuals \(N_r,\, F_r\) are shown in Figure~\ref{fig_nmc}c,d. We can see that for the entangled state transfer, deriving a new set of control fields tailored to the open-system dynamics generally performs better than just applying the closed-system controls to open systems. In~addition, the~memory effects are shown to be beneficial for the optimal state transfer and can achieve higher values than the Markov cases: the controlled non-Markovian system can reach smaller residuals than the Markov case, and~for the closed-system controls, they are more robust in non-Markovian scenarios and are penalized more under Markov~dynamics.

\begin{figure}[H]
 \includegraphics[width=.95\textwidth]{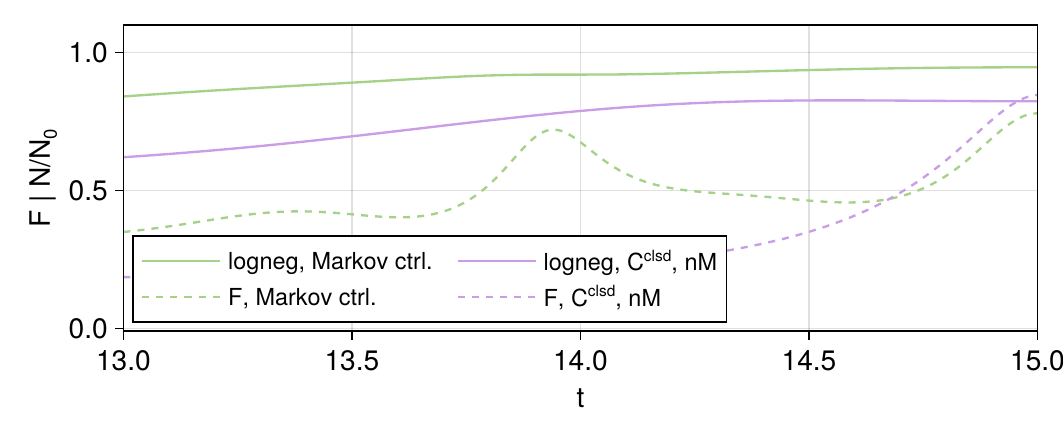}
 \caption{{Dynamics}  of the fidelity (dashed lines) and entanglement (solid lines). Compare the case where the control is calculated under the Markov open system (green lines) with applying the closed-system control in a non-Markovian setting (purple lines); we can see that while the non-Markovian case here has better fidelity than the Markov case at \(t=T\), it is actually less entangled, illustrating that a higher value of fidelity against the target entangled state does not necessarily mean a higher degree of entanglement is~achieved.}\label{fig_nmc_fidneg}
\end{figure}

One interesting observation here is that for the final residuals, calculating a new set of controls under Markov open-system dynamics gives a closer negativity than applying the close-system controls in a non-Markovian setting, while the latter have a higher fidelity. We show the dynamics of both cases near the end of the runtime \(T\) in Figure~\ref{fig_nmc_fidneg}. This clearly illustrates the need to choose the control optimization functional to target both fidelity and negativity since the overlap alone does not guarantee that the state can have a closer target entanglement, and~vice~versa, states with the same amount of entanglement can be far away in the Hilbert space or even have no overlap at all. One example in the discrete case would be the \(4\) Bell states, which are all maximally entangled but orthogonal to each~other.

\begin{figure}[H]
    \includegraphics[width=.9\textwidth]{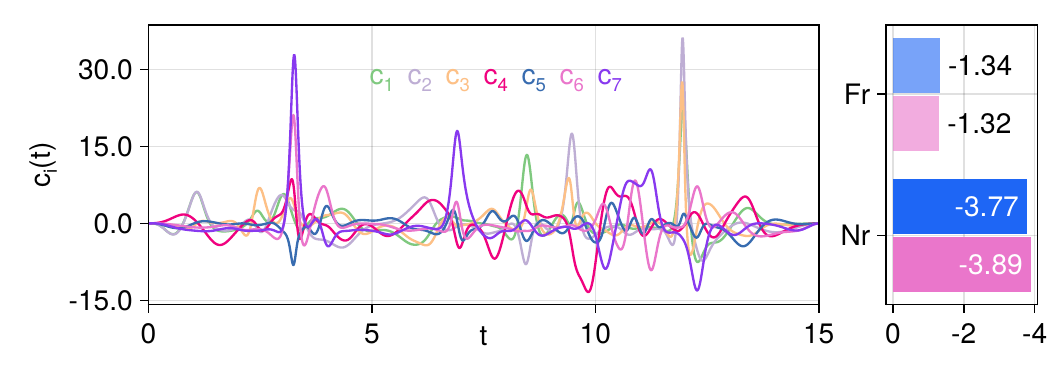}
    \caption{{(\textbf{Left}) panel:} Control fields obtained for the non-Markovian case but without the amplitude clamping. One can see that they can 
 reach high values with some sharp peaks, which may be challenging for experimental realizations. (\textbf{Right}) panel: Logarithmic of the residuals of fidelity \(F_r\) and negativity \(N_r\); pink colors are for the no-clamping case here, while the blue colors are for the clamped case of Figure~\ref{fig_nmc} for comparison. It can be seen that with the better-behaved clamped controls, similar residuals can still be~reached.}\label{fig_nm_noclamp}
\end{figure}

As a comparison, we show the control fields obtained from Krotov's methods without the amplitude clamping in Figure~\ref{fig_nm_noclamp}. We may observe the existence of some large values and sharp peaks, which may be challenging for experimental realization of the controls, whereas after the amplitude clamping via the \(\tanh\) function clamping, the~control fields in Figure~\ref{fig_nmc}a are smoother and much smaller in value and overall better-behaved.

\unskip

\begin{figure}[H]
 \includegraphics[width=.9\textwidth]{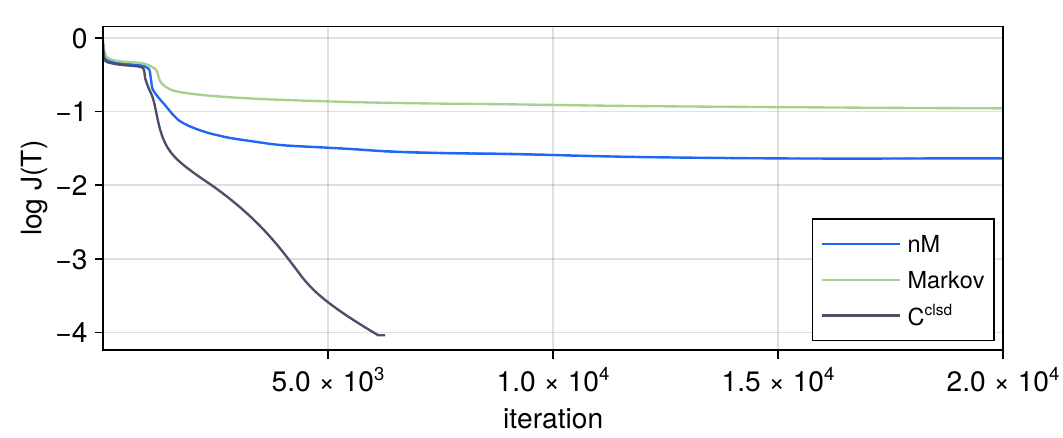}
 \caption{Logarithmic of the control function \(J(T)\) against the number of Krotov control iterations. It may be observed that the monotonic behavior is still being maintained with the modified \mbox{iteration~algorithm.}}\label{fig_nm_mono}
\end{figure}

One main strength of the Krotov control is that a monotonic convergence is guaranteed. Here, we have modified the iteration algorithm to update the dependence of the \(\bar{O}\)-operator on the control fields between iterations to obtain the correct effective Hamiltonian. With~an appropriately large inverse step-size \(\Lambda_{a,l}\) in Equation~(\ref{eq_ctrlup}) so that the field updates between iterations are small, we can still properly maintain the monotonic convergence properties. With~\(\Lambda_{a,l}=2\) for a closed system and  \(\Lambda_{a,l}=4\) for an open system, the~value of the control function Equation~\eqref{eq_ctrlj} against the number of iterations is shown in Figure~\ref{fig_nm_mono}. We can also see that for ideal closed systems, the~control target function quickly drops to very small values, whereas the descent of the open system is much slower and plateaus around \mbox{\(5000\)~iterations.}

\begin{figure}[H]
 \includegraphics[width=.95\textwidth]{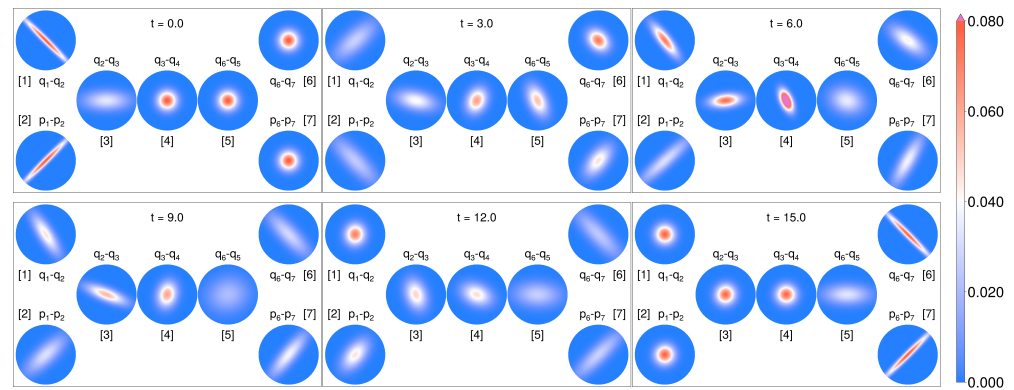}
 \caption{{Snapshots} of the Wigner functions for the controlled X-shaped chain under non-Markovian~dynamics.}\label{fig_wigner}
\end{figure}

To visualize how the entangled state progresses through the chain, we show the snapshots of the Wigner function in Figure~\ref{fig_wigner}. For~both ends, the~Wigner function is taken to be along the squeezed dimensions \(q_{1, N-1}-q_{2,N}\) and~\(p_{1, N-1}, p_{2,N}\), while the middle oscillators are set to neighboring~position--position.

\unskip

In practical scenarios, one may not know beforehand the details of the entangled state to transfer: for our example here, one may not know the squeezing parameter \(r\) of the TMSS. In~such a case, optimal control can still be made possible if we can assume that the parameter is in some broad range and~calculate \emph{{one}} set of control fields that can minimize the average of the control target functions for \emph{{some sample points}} of the states in some range. 
 As~an example, we take the linear chain and assume the squeezed parameter \(r\) is within \([0.6, 1.0]\). We then take \(5\) equally spaced sample points in this range to ``train'' one set of control fields that are effective for these sample points \emph{{simultaneously}}. This is possible because Krotov's method allows multiple pairs of initial/target states,
\begin{align}
|\Psi_i(r_j) \rangle &= S_{1,2}(r_j) |0 \rangle_{1\ldots N}, \\
|\Psi_T(r_j) \rangle &= S_{N-1,N}(r_j) |0 \rangle_{1\ldots N},
\end{align}
where \(r_j = 0.6, 0.7, \ldots, 1.0\). For~the average of the control functional, one may take the simple arithmetic average or~something else more suitable for the specific setup. Here, we want to put more weights on the most `problematic' parameter in the range so that the control update is more skewed to minimize the largest residuals. One average that fulfills this requirement is the log-sum-exp (\(\operatorname{LSE}\)) function
\begin{align}
  \operatorname{LSE}(x_1, \ldots, x_n) = \log\left(\sum_i e^{x_i}\right),
\end{align}
which can serve as a smooth approximation to the maximum of \(x_{1\ldots n}\). We also clamp the controls so that their amplitudes do not exceed \(A=10\).

\begin{figure}[H]
 \includegraphics[width=.9\textwidth]{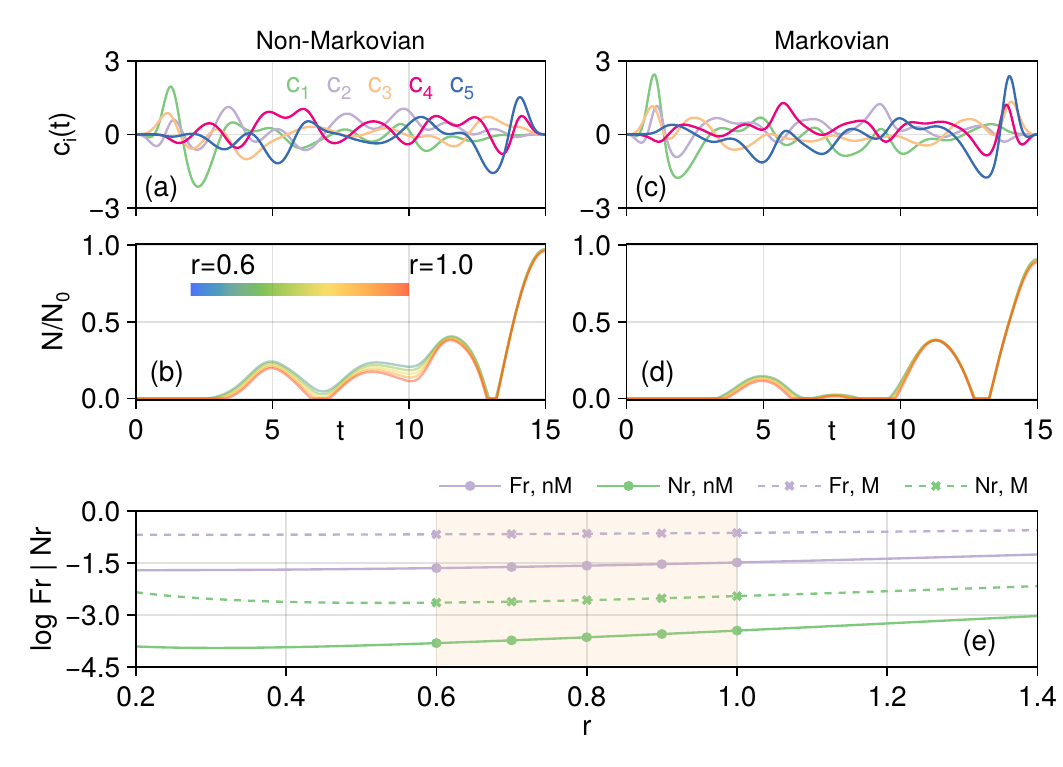}
 \caption{{Optimizing}  over a range of squeezing parameters for the linear chain under Markov and non-Markovian dynamics. By~targeting multiple initial--target state pairs in the squeezing parameter range \(r \in [0.6, 1.0]\), we can drive the desired entangled state transfer across a wide range of parameters. Panel (\textbf{a},\textbf{b}): Control fields as functions of time under non-Markovian noises and~negativity dynamics across different squeezing parameters. Panel (\textbf{c},\textbf{d}): the control and entanglement dynamics for the Markov case. Panel (\textbf{e}): Residuals of the control target across different squeezing parameters \(r\). The `training' ranges are orange-shaded, where the markers show the squeezing parameter values being considered. We can see that the control can work for parameters both inside and slightly outside of the range being targeted by the control, and~more importantly, non-Markovian memory effects can be beneficial for the optimal transfer of the entangled~state.}\label{fig_lch_multi}
\end{figure}

We show the resulting control fields and entanglement dynamics for both the non-Markovian and Markov cases in Figure~\ref{fig_lch_multi}. Krotov's method is set to run \mbox{20,000 iterations} for both cases with \(\Lambda_{a,l}=5\). For~the controls displayed in Figure~\ref{fig_lch_multi}a,c, we can see that the controls for the non-Markovian case are smaller in amplitude, while the resulting fidelity and entanglement are also much better than in the memoryless Markov case; the~improvement here is found to be around \(7\) to \(14\) times better. This shows the advantage of considering the non-Markovian memory effects, and~we show that it can aid in the optimal controlled transfer of entangled states in chains of harmonic~oscillators.

\section{Conclusions and~Discussion}

In this work, we investigated the optimal transfer of entanglement using Krotov's method in two different configurations of harmonic oscillator chains. We show that by tuning the individual oscillator's frequencies while keeping the coupling strengths fixed, it is possible to transfer entangled states, such as a two-mode squeezed state, from~the head of the chain to its end within a prescribed~runtime.

Our new optimal control scheme can account for both fidelity and an entanglement measure in the optimization functional. We demonstrate the necessity of including both quantities, as~fidelity alone, which represents the state overlap,  does not guarantee the desired level of entanglement. The~gradient of the control functional with respect to the evolved states can be efficiently computed numerically through automatic differentiation. We then extend the control framework to open-system settings, where the Krotov iteration must be adapted to incorporate the memory effects of non-Markovian dynamics. In~particular, the~dissipative terms in the leading-order master equation are governed by a time-dependent operator, in~contrast to the constant operator that appears in the Markovian case.
However, the~non-Markovian case requires correcting the dissipative term after the control field has been updated by Krotov's iteration. Using this revised control algorithm, we have shown that the transfer of entanglement can be realized in general open-system settings. Importantly, the~memory effects are shown to be beneficial to the controlled transfer of entangled states. For~potential experimental realizations, we have demonstrated that the amplitudes of the control may be clamped by using a scaled \(\tanh\) function, which adds an additional term to the control update equation by the chain rule. In~addition, we show that with appropriately chosen control parameters, the~monotonic convergence of Krotov's methods can be maintained even with the modified memory effect corrections.
We also considered the scenario in which the squeezing parameter of the initial state is not precisely known, but~only estimated within a certain range. In~this case, the~controlled state transfer can still be achieved by targeting several representative sample points within that range. Following this protocol, we find that the resulting control fields successfully drive the entangled state transfer both within and slightly beyond the assumed parameter range. Experimentally, such chains of coupled harmonic oscillators may be realized using a chain of coupled cavities~\cite{Buchmann2018a,Okamoto2013a}, atoms in microtraps~\cite{Barredo2016a}, or~superconducting quantum interference devices (SQUIDs) and circuit quantum electrodynamics systems as a chain of LC circuits~\cite{Blais2007a,Blais2021a}, where individual frequency tuning may be possible~\cite{Liao2010a}.

Finally, we note that the techniques developed here are generic and can be extended to other systems or control objectives where non-Markovian effects must be taken into account in the optimization of quantum dynamics. It would be of great interest to explore how the memory effects inherent in non-Markovian dynamics may be harnessed to enhance the performance of other types of quantum~technologies.
\vspace{6pt}

\authorcontributions{{Conceptualization,}
 T.Y.; methodology, T.Y. and D.-W.L.; software, D.-W.L.; validation, E.Y. and D.-W.L.; formal analysis, D.-W.L., E.Y., and T.Y.; investigation, D.-W.L., E.Y., and T.Y.; writing---original draft preparation, D.-W.L., E.Y., and T.Y.; writing---review and editing, D.-W.L., E.Y., and T.Y.; visualization, D.-W.L.; supervision, T.Y.; project administration, T.Y.; funding acquisition, T.Y. All authors have read and agreed to the published version of the~manuscript.}

\funding{{This} 
 research was funded by the U.S. Department of Defense (ACC-New Jersey under Contract No. W15QKN-24-C-0004).}

\dataavailability{{The raw data supporting the conclusions of this article will be made available by the authors on request.}}

\conflictsofinterest{The authors declare no conflicts of interest. The~funders had no role in the design of the study; in the collection, analyses, or~interpretation of data; in the writing of the manuscript; or in the decision to publish the~results.}

\appendixtitles{yes} 
\appendixstart
\appendix
\section[\appendixname~\thesection]{Homogeneous Linear Differential Equation for the Open-System CM} \label{appx_opencm_vec}
The equation of motion for the CM under open-system dynamics, Equation~\eqref{eq_nm_dcm} has one \(\gamma\)-independent diffusion term such that the column-stacked CM \(\vec{\gamma}\) follows
\begin{equation}
	\partial_t \vec{\gamma}(t) = \mathcal{L}_0(t) \vec{\gamma}(t) + \vec{d}(t),
\end{equation}
which does not strictly follow the homogeneous differential equation form that Krotov's method takes. Nevertheless, we may pad a constant element \(\vec{\gamma}_1 = [\gamma, 1]\), which does follow a homogeneous equation\vspace{12pt}
\begin{align}
	\partial_t \vec{\gamma_1}(t)
  &= \partial_t \begin{pmatrix} \vec{\gamma} \\ 1 \end{pmatrix}
  = \begin{bmatrix} \mathcal{L}_0(t) && \vec{d}(t) \\ 0 && 0 \end{bmatrix} \begin{pmatrix} \vec{\gamma} \\ 1 \end{pmatrix} \\
  &\equiv \mathcal{L}_{\rm eff}(t) \vec{\gamma}_1(t).
\end{align}

\section[\appendixname~\thesection]{Control Field Update in Non-Markovian Open Systems: O-Operator Contribution to the Update Equation} \label{appx_krotovo}
Krotov's control update Equation~\eqref{eq_ctrlup} requires the derivative of the control function with respect to the effective Hamiltonian for Schr\"odinger-like equations. At~first glance, given the dependence of the \(O\)-operator on the control field, there would be some complex contribution to the control field from the functional derivative of the \(O\)-operator with respect to the control fields. While the \(O\)-operator follows a non-linear differential equation, so analytical solutions or numerical derivative can be challenging, we can show that the contribution from the \(O\)-operators to the control update equation may be approximately omitted: taking Equation~\eqref{eq_dodt}, we have
\begin{align}
  \bar{O}^{(0)}(t) &\approx \bar{O}^{(0)}(t-dt) + dt \partial_t \bar{O}^{(0)}(t-dt) \nonumber \\
  &= \bar{O}^{(0)}(t-dt) + \alpha(0)Ldt - \xi_{\rm eff}\bar{O}^{(0)}(t-dt)dt  \nonumber \\
  &+ \left[-iH_s(t-dt) - L ^\dagger \bar{O}^{(0)}(t-dt), \bar{O}^{(0)}(t-dt)\right] dt,
\end{align}
such that on a fine time grid, the~derivative \(\partial/\partial c_i(t)\) on the right-hand side may be approximately~omitted.


\begin{adjustwidth}{-\extralength}{0cm}

 \reftitle{References}



 \PublishersNote{}
\end{adjustwidth}
\end{document}